%% file: asplos24-paper-template.tex
\documentclass[nonacm,sigplan]{acmart}

\usepackage{braket}
\usepackage{savesym}
\usepackage{subcaption}
\usepackage{graphicx}
\usepackage{blindtext}
\usepackage{mathtools}
\usepackage{algorithm}
\usepackage[noend]{algorithmic}
\savesymbol{COMMENT}
\usepackage{lipsum}
\usepackage{graphicx}
\usepackage{color}
\usepackage{amsmath}   
\usepackage{amsthm}
\usepackage{amsfonts}
\usepackage{wrapfig}

\newcommand{\asplossubmissionnumber}{XXX}

\usepackage[normalem]{ulem}

\input{macros}

\newcommand{\squishlist}{
    \begin{list}{$\bullet$}
        { \setlength{\itemsep}{0pt}      \setlength{\parsep}{0pt}
            \setlength{\topsep}{0.5pt}       \setlength{\partopsep}{0pt}
            \setlength{\listparindent}{-2pt}
            \setlength{\itemindent}{-5pt}
            \setlength{\leftmargin}{0.5em} \setlength{\labelwidth}{0em}
            \setlength{\labelsep}{0.2em} } }
    
\newcommand{\squishend}{
\end{list}  }

\newtheorem{defi}{Definition}

\begin{document}

\title{
Minimizing Photonic Cluster State Depth in Measurement-Based Quantum Computing}

\author{Yingheng Li,  Aditya Pawar,
Zewei Mo, Youtao Zhang, Jun Yang, 
Xulong Tang}
\affiliation{
\institution{University of Pittsburgh}
\city{Pittsburgh}
\country{United States}
}

\date{}

\begin{abstract}
Measurement-based quantum computing (MBQC) is a promising quantum computing paradigm that performs computation through ``one-way'' measurements on entangled quantum qubits. It is widely used in photonic quantum computing (PQC), where the computation is carried out on photonic cluster states (i.e., a 2-D mesh of entangled photons). In MBQC-based PQC, the cluster state depth (i.e., the length of one-way measurements) to execute a quantum circuit plays an important role in the overall execution time and error. Thus, it is important to reduce the cluster state depth. 
In this paper, we propose FMCC, a compilation framework that employs dynamic programming to efficiently minimize the cluster state depth.  Experimental results on five representative quantum algorithms show that FMCC achieves 53.6\%, 60.6\%, and 60.0\% average depth reductions in small, medium, and large qubit counts compared to the state-of-the-art MBQC compilation frameworks. 
\end{abstract}

\maketitle 
\pagestyle{plain}

\section{Introduction} 
\label{sec:intro}

Quantum computing has experienced rapid development in the past decades, demonstrating its supremacy over classical computers in various application fields~\cite{shor,grover,chemistry}. 
Most quantum computing primarily focuses on matter-based quantum computers that utilize superconducting \cite{super} and trapped-ion \cite{ion} qubits, following the gate-based quantum computation paradigm. 
Recently, there has been a growing interest in photonic quantum computing (PQC) that employs photon qubits~\cite{mbqc,optical,photonicreview}. PQC generally leverages measurement-based quantum computation (MBQC) that carries out computation through ``one-way'' measurements on a 2-D mesh of entangled photon qubits (known as cluster states)~\cite{mbqc,mbqccluster}. Specifically, computation is conducted by measuring the photons and propagating the measurement results through a cluster state~\cite{mbqccluster, mbqc, fly}. MBQC-based PQC offers several advantages compared to gate-based quantum computation, such as longer coherence times and better scalability~\cite{photonic}. 

In MBQC, a quantum algorithm is executed by mapping the measurement patterns of each quantum gate onto a cluster state (detailed discussion in Section~\ref{sec:cluster}). A cluster state is generated deterministically using cluster state generator hardware that produces entangled photons with fixed cluster state width, where the width is the number of rows of entangled photons in the 2-D mesh~\cite{mbqc, mbqccluster,coupleemitter1, coupleemitter2,pichler2017universal,24row,2023deterministic}. In MBQC, the computation is realized by measurements to process and transmit information between neighboring entangled photons. As such, the depth (i.e., the number of photon columns to execute a quantum algorithm) is a metric that critically affects the execution time of quantum algorithms. 

In this paper, we observe that modern MBQC execution on cluster states is not optimized for the cluster state depth, resulting in a long depth that induces a large photon resource cost ~\cite{coupleemitter1,coupleemitter2,pichler2017universal} and increases quantum error rate~\cite{quantumdot, photonicreview, detector1}. To this end, we propose FMCC (\underline{F}lexible \underline{M}BQC \underline{C}ompilation on \underline{C}luster State), an MBQC compilation framework that achieves the minimized depth to execute a quantum algorithm on a cluster state. Specifically, the motivation stems from our observations that i) a significant amount of photons are wasted in the cluster state during execution, and ii) the MBQC execution allows different valid mapping ``variants'' that can leverage the wasted photons to reduce the depth. To generate different MBQC mapping variants that utilize the wasted photons, we categorize the measurements in an MBQC circuit into ``components'' --- each being a group of neighboring measurements. We identify two types of mapping variants, intra-component and inter-component, and employ dynamic programming to exploit the variants to generate the MBQC circuit with minimized depth on a cluster state. 

To the best of our knowledge, the only existing work that optimizes the depth is proposed by Li et al.~\cite{dac}. While their work primarily focuses on reducing the number of redundant photon measurements, they do not explore MBQC mapping variants, yielding sub-optimal depth. In this paper, we use their work as our baseline and provide a quantitative comparison in Section~\ref{sec:single} and Section~\ref{sec:multi-round}. The main contributions of the paper are the following:

\squishlist{}
    \item We identify cluster state depth as a key metric for reducing the execution time of quantum algorithms on photonic cluster states. We find that an alternative MBQC mapping effectively reduces cluster state depth.
    
    \item We categorize the photon measurements on cluster state into different components. We define the component constraints in circuit mapping and use components to generate mapping variants. Then, we formulate the problem into a dynamic programming problem and exploit both intra-component and inter-component mapping variants. This significantly reduces the mapping search space and avoids simple heuristic pitfalls. 
    
    \item We conduct experiments using five representative MBQC quantum applications with a small, medium, and large qubit count. Experimental results indicate that, compared to the state-of-the-art MBQC compilation, FMCC achieves 53.6\%, 60.6\%, and 60.0\% average depth reductions in small, medium, and large qubit counts, respectively.
\squishend{}


\begin{figure}
    \centering
    \includegraphics[width=1\linewidth]{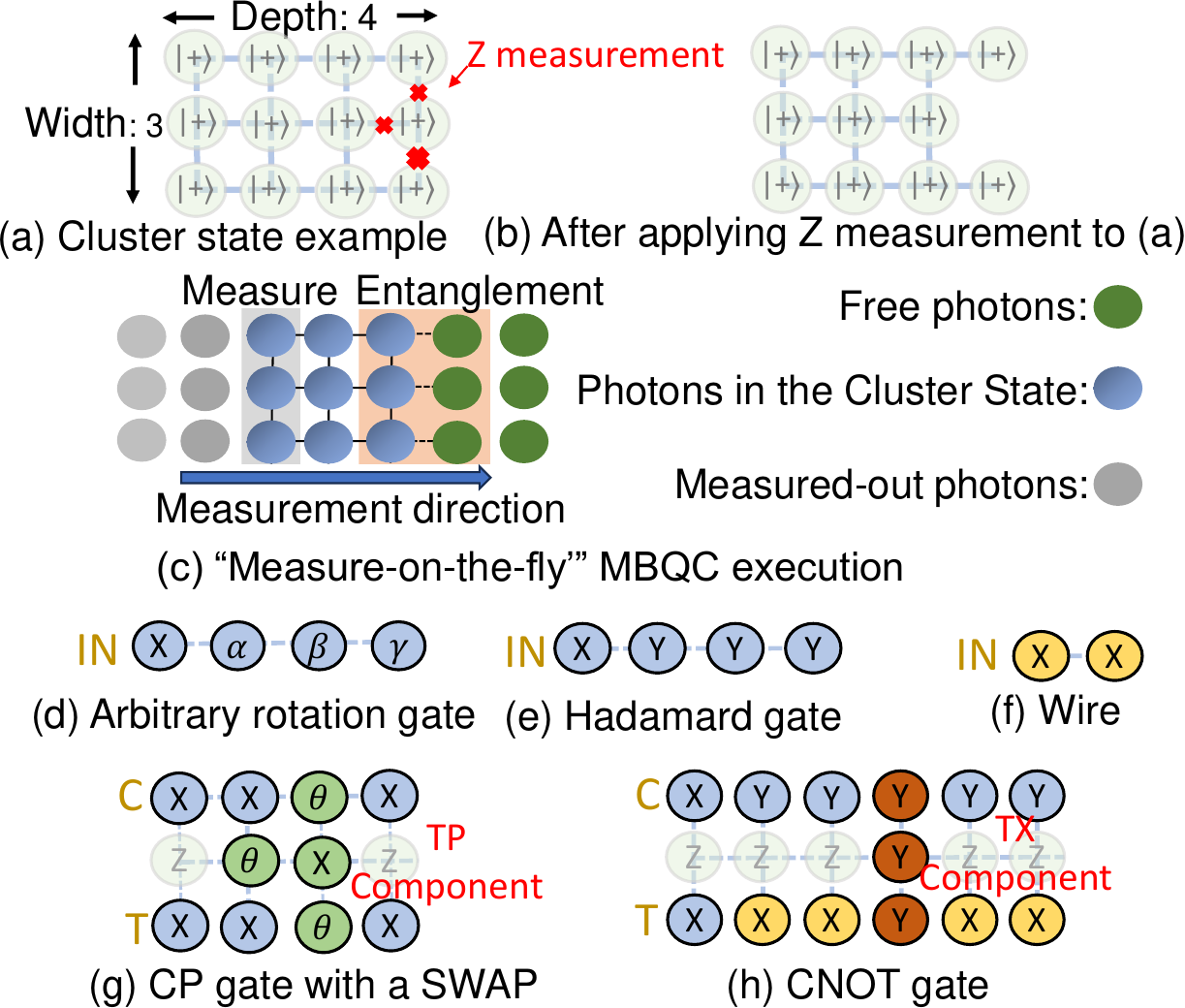}
    \caption{Cluster state and Measurement pattern examples.}
    \label{fig:gates}
\end{figure}

\section{Background}
\label{sec:background}

Measurement-based quantum computing (MBQC) carries out the computation on a 2-D mesh of entangled photon qubits, which is referred to as a {\it cluster state}. A cluster state consists of photon qubits initially prepared in the $\ket{+}$ state where each photon is entangled with its neighboring photons. Fig.~\ref{fig:gates}(a) illustrates a photonic cluster state with 12 photons. We next elaborate on the details of photonic cluster states (Section~\ref{sec:cluster}) and MBQC on cluster states (Section~\ref{sec:mbqc basis}).

\subsection{Photonic Cluster State}
\label{sec:cluster}
A cluster state is generated by a photonic hardware generator (i.e., cluster state generator) that continuously produces entangled photons~\cite{coupleemitter1, coupleemitter2, coupleemitter3, pichler2017universal, 24row, 2023deterministic, quantumdot}. Fig.~\ref{fig:gates}(c) shows such a generation process in which the generator has a fixed width of 3 photons and the entangled photons are generated from left to right. In the 2-D mesh cluster state, the {\it width} and {\it depth} are important for a quantum circuit execution on a cluster state~\cite{mbqccluster,dac,introduction}. 

{\bf Cluster state width.} The width refers to the number of rows in a cluster state. A fixed width simplifies photonic cluster state generator hardware, which is a typical design~\cite{coupleemitter1, coupleemitter2, coupleemitter3, pichler2017universal, 2023deterministic, 24row}. However, the minimum width required by each quantum algorithm varies according to the number of logical qubits it has~\cite{mbqccluster, dac, mbqc}. Specifically, each logical qubit is mapped to a row of photons in the cluster state. To prevent undesired interactions between two logical qubits, the corresponding photon rows cannot be adjacent since the adjacent photons are entangled in a cluster state. Therefore, MBQC cuts out one row of photons between the two rows where logical qubits are mapped. Hence, a quantum algorithm requires a minimum of $2N_q-1$ rows to be executed in a cluster state, where $N_q$ represents the number of qubits in the quantum algorithm~\cite{mbqccluster, fly, dac}. 


{\bf Cluster state depth.} The depth refers to the number of columns in a cluster state. A hardware generator produces entangled photons from left to right, where there are no hardware generator constraints in the cluster state depth. Thus, the depth of the cluster state is only determined by a quantum algorithm and its mapping on the cluster state. The depth is necessary to reduce execution time: the shorter the cluster state depth, the faster the execution and the lower the error rate~\cite{coupleemitter1, coupleemitter2, coupleemitter3, pichler2017universal, 2023deterministic, quantumdot, photonicreview}. Therefore, our goal is to reduce the depth in a fixed-width cluster state to execute a quantum algorithm.

Fig.~\ref{fig:gates}(c) also illustrates the ``measure-on-the-fly'' execution paradigm employed in the cluster state~\cite{fly,photonicreview,optical}. That is, the measurements are applied to the photons while the cluster state is generated. This avoids the need to store a large-scale photonic cluster state in optical quantum memory, increasing MBQC's scalability~\cite{fly,photonicreview,optical}. Note that the measurement direction must be the same as the cluster state generating direction in this paradigm (i.e., from left to right). As shown in Fig.~\ref{fig:gates}(c), the cluster state generator entangles free photons (green circles) into a cluster state (blue circles). Meanwhile, a column of photon detectors measures the leftmost column of the cluster state to perform computations. Once measured, the photons (gray circles) are no longer entangled. 


\begin{figure*}
    \centerline{\includegraphics[width=0.95\linewidth]{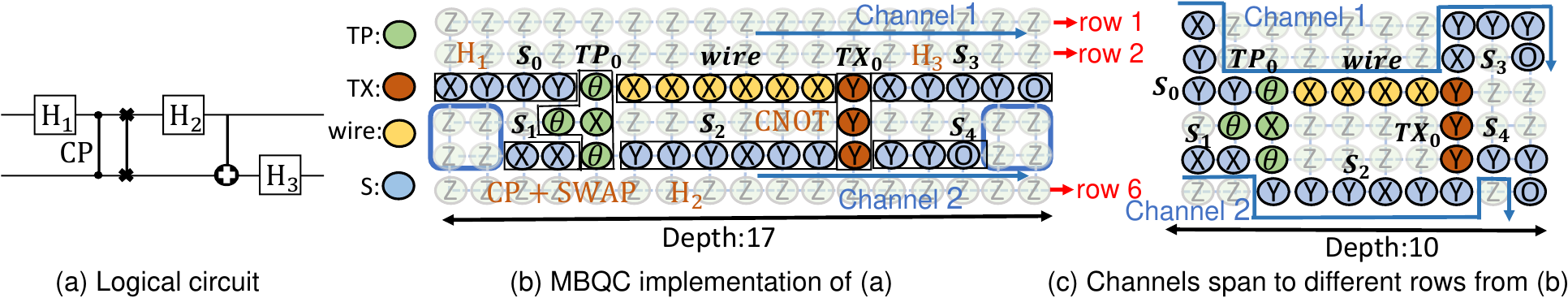}}
    \caption{Example of a quantum algorithm mapped on a cluster state.}
    \label{fig:cluster}
\end{figure*}

\subsection{MBQC on Cluster State} 
\label{sec:mbqc basis}

MBQC employs two types of measurements to achieve universal quantum computation~\cite{oneway, mbqccluster}: i) X and Y measurements ($\sigma_x$ and $\sigma_y$ eigenbasis) to perform Pauli gate operations~\cite{oneway, mbqccluster}, and ii) $\theta$ measurements to perform non-Pauli gate operations. Implementation of quantum gates (also called ``pattern''~\cite{mbqccluster}) is achieved by a set of measurements applied to entangled photons in cluster states. Fig.~\ref{fig:gates} shows several pattern examples with annotated measurement angles. The patterns of a single-qubit arbitrary rotation gate and an H gate are shown in Fig.~\ref{fig:gates}(d) and (e), where measurements are applied to a row of consecutive entangled photons and computation is carried out ``one-way'' from left to right. The consecutive X measurements of even numbers serve as identity gates or ``wires'', allowing logical qubit transmission through qubit rows without applying any quantum operations~\cite{mbqccluster} (Fig.~\ref{fig:gates}(f)).   

MBQC uses Z measurements to ``cut out'' entangled photons from a cluster state~\cite{oneway, mbqccluster}. Fig.~\ref{fig:gates}(a) shows an example of applying a Z measurement to a photon in the cluster state, destroying the entanglements between the photon and its neighbors. The resulting cluster state is shown in Fig.~\ref{fig:gates}(b). The Z measurements eliminate unwanted interactions between logical qubits by separating qubit rows in a cluster state. It also eliminates any photons within the cluster state that are not involved in the computation.

Two notable two-qubit gate patterns are the combined CP and SWAP gate and the CNOT gate, shown in Fig.~\ref{fig:gates}(g) and (h), where measurements are applied to three rows of entangled photons. The first row corresponds to the control qubit, whereas the third row corresponds to the target qubit. Among the measurements in the middle row, the X, $\theta$, and Y measurements facilitate communication between the two logical qubits, while the Z measurements cut out unused photons and separate rows for control and target qubits. While CNOT and single-qubit gates together enable universal quantum computation~\cite{universality,nielsen2001quantum}, MBQC provides further advantages as CNOT gates together with combined CP and SWAP gates reduce the need for quantum gate decomposition and enhance the flexibility of quantum computation~\cite{mbqccluster}.




\section{Opportunities and Challenges}
\label{sec:motivate}

We use a real MBQC example to illustrate the opportunities and challenges in optimizing the cluster state depth. Fig.~\ref{fig:cluster}(a) shows a two-qubit quantum algorithm, where its corresponding MBQC implementation on a photonic cluster state is shown in Fig.~\ref{fig:cluster}(b). 
After the computation on a logical qubit (i.e., a photon row), an output photon is employed for readout measurement $O$. Fig.~\ref{fig:cluster}(b) shows the mapping from a quantum algorithm to a cluster state, which we refer to as an MBQC circuit in the rest of this paper. 

The one-way computation in MBQC can result in long cluster state depth, which is often due to inefficient photon usage. For example, only three of six photon rows are being used in Fig.~\ref{fig:cluster}(b). That is, rows 1, 2, and 6 in Fig.~\ref{fig:cluster}(b) are not used by effective computation and are cut out by Z measurements. Moreover, there are Z measurements that cut out photons \textit{within} the three used photon rows, as indicated by the blue boxes. This motivates the question: \textit{Can we leverage these unused photons and rows to reduce the circuit depth?}

\subsection{MBQC Circuit Variants}
\label{sec: opportu}
In practice, the measurements corresponding to a logical qubit in a cluster state are not necessarily mapped to the same row. This is because the quantum computation on a logical qubit remains unchanged as long as the same measurements are made on entangled photons in the same order as in a cluster state~\cite{oneway,mbqccluster}. That is, one can leverage the entangled photons in the same column, rather than being strictly limited to entangled photons in the same row. We define the measurement sequences on entangled photons in cluster states as ``channels''. 
Note that measurements within a column can be performed either simultaneously or sequentially~\cite{photonicreview,optical}. In the baseline MBQC circuit mapping, one channel is necessarily mapped to one photon row. However, if we allow the channel to span multiple rows, it can leverage unused photons and potentially reduce depth.

Different MBQC circuits can be generated when the channel spans multiple photon rows in a cluster state.
To explore shorter MBQC circuits with the same function, we first partition an MBQC circuit into different components, then explore the variant of each component and connect them together. Each component is a group of adjacent measurements with a specific functionality. We define
four types of components:

\squishlist{}
    \item \textbf{TP}: The TP component is used to connect the control qubit (i.e., control channel) and the target qubit (i.e., target channel) in the combined CP and SWAP gate pattern. It consists of one X and three $\theta$ measurements. The four measurements of a TP component are colored in green in Fig.~\ref{fig:gates}(g).

    \item  \textbf{TX}: The TX component consists of three Y measurements and is used to connect the control qubit and the target qubit in CNOT gate patterns. The TX component is colored in red in Fig.~\ref{fig:gates}(h). The three Y measurements in the TX component are required to be measured simultaneously~\cite{mbqccluster}.

    \item \textbf{S}: The S component represents single-qubit measurements. An S component consists of consecutive measurements within individual channels, e.g., Fig.~\ref{fig:gates}(d) and Fig.~\ref{fig:gates}(e). Note that an S component cannot include pairs of X measurements because they only form wires.
   
   \item\textbf{wire}: A wire is a consecutive sequence of an even number of X measurements. As previously discussed, a wire only serves as a connector between two components and does not perform any quantum computation.
    
\squishend{}
  
  

Based on these four types of components, we next define intra-component variants and inter-component variants.   

\begin{figure}
    \centering
   \includegraphics[width=1\linewidth]{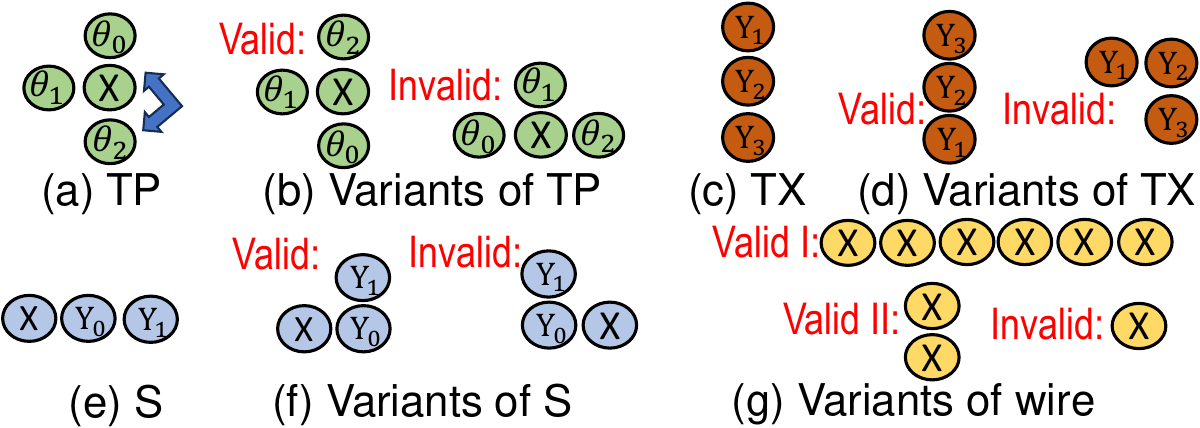}
    \caption{Intra-component variants.}
    \label{fig:variants}
 \end{figure}

{\bf Intra-component variants: }For each component, there are multiple MBQC mappings that have the same quantum functionality as in a baseline MBQC circuit. Specifically, each type of component has its own unique set of variants:   
\squishlist{}
  \item \textbf{TP variants}: A TP component has only one valid variant. Fig.~\ref{fig:variants}(a) shows the TP component in the baseline MBQC circuit, whereas Fig.~\ref{fig:variants}(b) shows its valid variant and one example of an invalid variant. The valid variant swaps the $\theta_0$ and $\theta_2$ measurements in the first and third rows, respectively.
  As discussed in Section~\ref{sec:cluster}, the measurements in a cluster state are performed from left to right. This implies that measurements in the first column should be executed before those in the second column. As such, the valid variant preserves the same order as the baseline because the same measurements are present in each column. In contrast, the invalid variant does not preserve the same measurement order, resulting in an incorrect computation due to a different measurement sequence~\cite{oneway,mbqccluster}.   
  
  \item \textbf{TX variants}: Similar to the TP component, the TX component also has only one valid variant. Fig.~\ref{fig:variants}(c) shows the baseline TX component and Fig.~\ref{fig:variants}(d) shows its valid variant and one example of an invalid variant. The valid TX variant also swaps the baseline TP component measurements. In both the baseline TX component and the valid TX variant, the three Y measurements can be performed simultaneously as they are in the same photon column. However, in the invalid variant, Y$_2$ and Y$_3$ are measured after Y$_1$, leading to incorrect computation. 
  
  \item \textbf{S variants}: Fig.~\ref{fig:variants}(e) shows an example of an S component with three measurements. In the baseline S component, measurements follow a left-to-right sequence, from X to Y$_1$. Although the two Y measurements are in the same column in the valid variant shown in Fig.~\ref{fig:variants}(f), X is measured first in the left column and Y$_0$ can be measured before Y$_1$~\cite{photonicreview,optical}. Hence, the same measurement order can still be kept. In contrast, in the invalid variant, the X measurement is to the right of the two Y measurements and it would be measured after these two.
  
  \item\textbf{wire variants}: A wire component only requires an even number of X measurements. Fig.~\ref{fig:variants}(g) shows three variants of a wire component. The valid variant I has six X measurements and the valid variant II has two. The invalid variant has one measurement, changing the wire's function.   
\squishend{}

\begin{figure}
    \centering
    \includegraphics[width=1\linewidth]{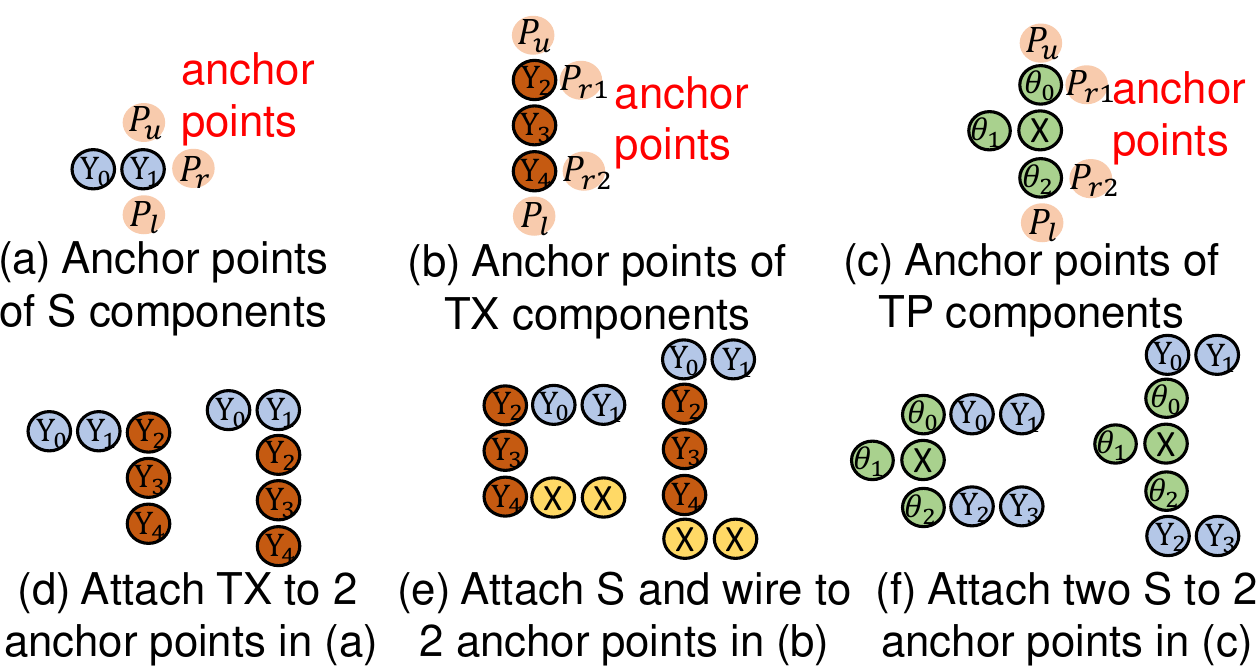}
    \caption{Anchor points in components.}
    \label{fig:positions}
\end{figure}

\noindent{\bf Inter-component variants: } 
Between two adjacent components, we refer to the component on the left as the ``left component'' and the component on the right as the ``right component,'' and refer to measurements linking these adjacent components as ``connected points.'' By adjusting the positions between two adjacent components, we can decrease the depth of an MBQC circuit by optimizing photon use in a cluster state. In this process, it is essential to maintain the connection between adjacent components, where the right component's connected point should attach to the ``anchor points'' on the left component. Anchor points are defined as the unused neighboring photons of the left component's connected point, where the right components can be attached. For example, Fig.~\ref{fig:positions}(a) shows the three anchor points of the S component ($P = \{P_u, P_r, P_l\}$). Applying the TX component to $P_u$ is not possible as it results in overlapping measurements, which is not allowed in MBQC~\cite{mbqccluster}. As shown in Fig.~\ref{fig:positions}(d), anchoring at $P_l$ reduces circuit depth compared to $P_r$. Unlike the S component, TP and TX components offer two anchor points, as shown in Fig.~\ref{fig:positions}(b) and (c). In Fig.~\ref{fig:positions}(e) and (f), each right component of the TP and TX components applies to a different pair of anchor points, where the pair $(P_u, P_l)$ minimizes circuit depth. 

\noindent{\bf Illustrating intra- and inter-component opportunities: }Exploiting both intra-component and inter-component variants allows us to reduce the circuit depth while maintaining the correct quantum function. An example is shown in Fig.~\ref{fig:cluster}(c) which is functionally equivalent to the MBQC circuit in Fig.~\ref{fig:cluster}(b). Specifically, for intra-component, we use variants of S$_0$, S$_3$, and S$_4$ that span more rows to reduce the depth. For inter-component, we attach S$_2$ to the lower anchor point of TP$_0$, attach TX$_0$ to the upper anchor point of S$_2$, and attach S$_3$ to the upper anchor point of TX$_0$. As a result, the depth is reduced from 17 (in Fig.~\ref{fig:cluster}(b)) to 10 (in Fig.~\ref{fig:cluster}(c)) by leveraging both variants.


\subsection{MBQC Constraints}
\label{sec:constraints}

While exploiting intra- and inter-component variants allows us to reduce the depth, it is a nontrivial task where certain constraints have to be satisfied to guarantee correctness.  

\begin{figure}
    \centering
    \includegraphics[width=1\linewidth]{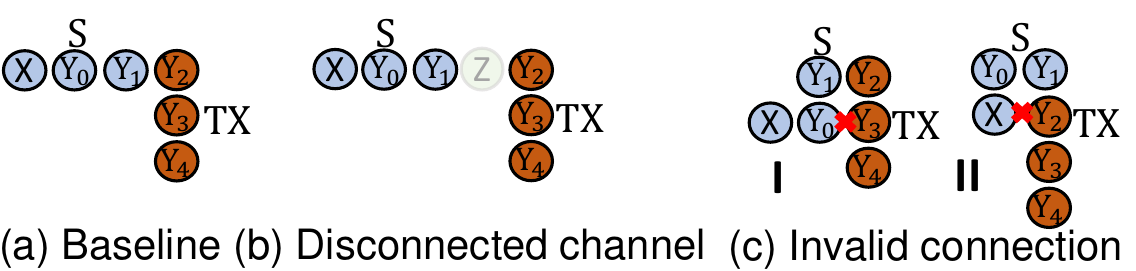}
    \caption{Examples of violating constraints.}
    \label{fig:constraints}
\end{figure}

\noindent{\bf Constraint-I Channel connectivity.} It is crucial to ensure channel connectivity in the MBQC circuit. That is, the sequence of measurements (i.e. the channel) should remain the same as in the baseline circuit. This constraint is violated if a channel is disconnected between two adjacent measurements. In Fig.~\ref{fig:constraints}(a), we show an S component adjacent to a TX component. Fig.~\ref{fig:constraints}(b) shows this constraint violation due to a disconnected channel.

\noindent{\bf Constraint-II Invalid connection.} Measurements that have no connection in the baseline MBQC circuit should remain disconnected. This violation occurs when using an inappropriate variant or connecting components to inappropriate anchor points. For instance,
in example I of Fig.~\ref{fig:constraints}(c), an invalid connection, marked by the red cross, is caused by the adoption of an inappropriate variant of the S component where the Y$_0$ and Y$_3$ measurements are connected. In example II of Fig.~\ref{fig:constraints}(c), there is an invalid connection between X and Y$_2$ due to the inappropriate anchor points.

\noindent{\bf Constraint-III Component dependency.} The dependency among the components has to be maintained when exploring MBQC circuit variants to ensure quantum computation correctness. That is, any component (including all the measurements in it) should not be conducted before those components it depends on. 


\noindent{\bf Constraint-IV Cluster state width. } The depth-reduced MBQC circuit must not have variants that exceed the fixed width of the cluster state. Violating this constraint makes the quantum circuit inexecutable on the cluster state~\cite{mbqccluster, dac, mbqc}.

\section{Our Approach}
\label{sec:approach}

In this paper, we design FMCC (\underline{F}lexible \underline{M}BQC \underline{C}ompilation on \underline{C}luster State), an MBQC compilation framework that exploits both intra-component and inter-component variants while satisfying the aforementioned constraints. However, this process faces two challenges. First, the design space can be prohibitively large, making exhaustive search practically infeasible. For instance, given a circuit with $n$ components, if each component has $v$ variants and $k$ anchor points, conceptually, one can have $v^n \times k^{n-1}$ MBQC circuits. While a simple heuristic may work, the second challenge is that among the $v^n \times k^{n-1}$ MBQC circuits, many of them may be invalid due to constraints violations. As such, a heuristic approach may lead to invalid MBQC circuits. To this end, our proposed FMCC employs dynamic programming (DP) to construct a valid MBQC circuit with reduced depth, while addressing the two challenges.


\begin{figure}
    \centering
    \includegraphics[width=0.9\linewidth]{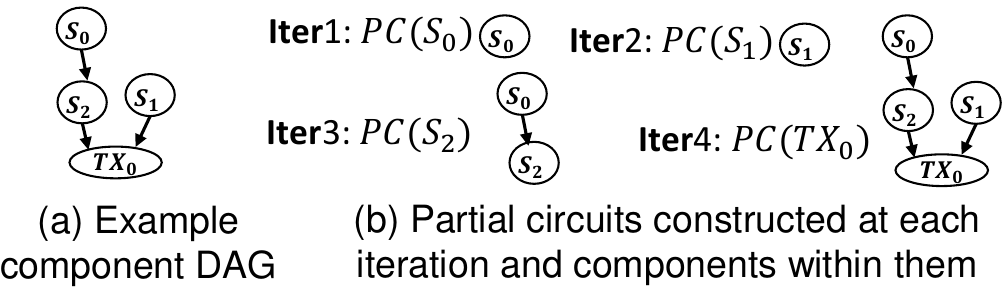}
    \caption{Partial circuit set constructed at each iteration.}
    \label{fig:pc}
\end{figure}

\subsection{Constructing MBQC Circuits}
\label{sec: DAG}

In FMCC, MBQC circuits are built component-wise. The construction sequence is guided by a Directed Acyclic Graph (DAG), which captures the dependencies among non-wire components, referred to as a component DAG\footnote{The component dependency is inherited from the gate dependency in the original logic circuit. Capturing the gate dependency has been substantially explored by prior quantum compilers~\cite{gushu,dac,jiliu}.}. We iterate through each component of the DAG, to generate a set of variants for that component. In each iteration, one component is chosen from the DAG. We define a variant set for a component chosen at iteration $k$ as:  

\begin{defi}(Variant set)
    The variant set $V(c)$ at iteration $k$ is the set of all valid variants of the component $c$ selected at iteration $k$, including its baseline variant.
\end{defi}
Starting from the independent components, i.e., components without parents in the DAG, a set of partial circuits is constructed iteratively by integrating the set of variants for the selected component. We define the partial circuit set as: 


\begin{defi}(Partial circuit set)
    The set of partial circuits $PC(c_i)$ for the selected component $c_i,$ where $i \in \{1, ..., n\}$ for $n$ components in the DAG, has elements constructed using the set of partial circuits for $c_{i-1}$ (i.e., $PC(c_{i-1})$) with the variant set of $c_i$ (i.e., $V(c_i)$), where $c_{i-1}$ is the parent component of component $c_i$ in the DAG. That is,  
    \begin{gather}
    \label{eq:1}
        PC(c_i) = \cup_{z \in P} (PC_x \oplus V_y) \\ 
        \forall\ PC_x \in PC(c_{i-1}),\ and\ \forall\ V_y\ \in\ V(c_i) \nonumber
    \end{gather}
    where $PC_x$ is a partial circuit in $PC(c_{i-1})$, $V_y$ is a variant of component $c_i$ in $V(c_i)$, $\oplus$ stands for both component integration and partial circuit combination, and $P$ is the set of all anchor points in $PC_x$. 
\end{defi}

To illustrate the partial circuit set constructed in each iteration, consider the component DAG in Fig.~\ref{fig:pc}(a) which includes $n=4$ components. In the first iteration, the independent component S$_0$ is selected and all its valid variants are generated as $V(S_0)$. Since S$_0$ lacks parent components, partial circuit set $PC(S_0)$ is generated using the initial partial circuit set $\emptyset$ and $V(S_0)$. Similarly, in the second iteration, $PC(S_1)$ is generated using $\emptyset$ and $V(S_1)$. In the third iteration, $PC(S_2)$ is generated by integrating $PC(S_0)$ and $V(S_2)$ since S$_2$ has a parent component S$_0$ in the DAG. In the fourth iteration,  $PC(TX_0)$ is generated by integrating $PC(S_2)$, $PC(S_1)$, and $V(TX_0)$ according to Equation~\ref{eq:1}.


With the definitions of variant and partial circuits, we next elaborate on how FMCC constructs an MBQC circuit. At each iteration, one independent component is selected and removed from the component DAG. Based on the position of the component in the DAG, FMCC performs either component integration or partial circuit combination. Following this, wires may be inserted.

\begin{figure}
    \centering
    \includegraphics[width=1\linewidth]{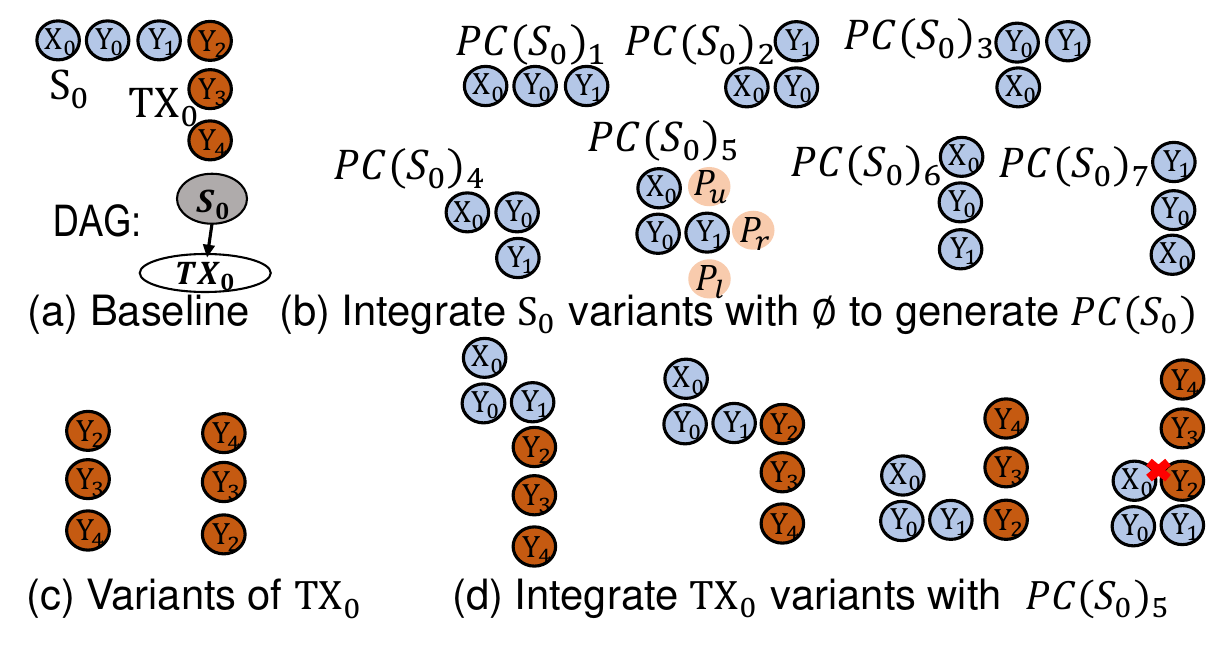}
    \caption{Example of component integration.}
    \label{fig:initial}
\end{figure}

{\bf Component integration.} 
At the beginning, FMCC starts with the initial partial circuit set as $\emptyset$ and selects the first independent component from the DAG. If there are multiple independent components, FMCC randomly chooses one of them.  Fig.~\ref{fig:initial}(a) shows an example DAG with two components S$_0$ and TX$_0$. Since S$_0$ is independent, FMCC chooses S$_0$ as the first component. Next, FMCC generates the variant set of S$_0$, i.e. $V(S_0)$, which contains all the variants of S$_0$, as shown in Fig.~\ref{fig:initial}(b). Based on equation~\ref{eq:1}, the partial circuit set $PC(S_0)$ is constructed by adding $V(S_0)$ to $\emptyset$, the initial partial circuit set. Next, TX$_0$ becomes independent as S$_0$ is removed from the DAG in the previous iteration. FMCC chooses TX$_0$ and generates its variants set $V(TX_0)$ as shown in Fig.~\ref{fig:initial}(c). Then, FMCC performs component integration using the partial circuit set $PC(S_0)$ and the variants set $V(TX_0)$. According to equation~\ref{eq:1}, all the anchor points of each partial circuit $PC_x$ in $PC(S_0)$ are explored with each variant $V_y$ in $V(TX_0)$. Take as an example the 5th partial circuit ($PC(S_0)_5$ in Fig.~\ref{fig:initial}(b)) and the 2nd variant of TX$_0$, Fig.~\ref{fig:initial}(b) shows the three anchor points ($P = \{P_u, P_r, P_l\}$) of $PC_5$. FMCC integrates the TX$_0$ variants to each of the anchor points, generating partial circuits shown in Fig.~\ref{fig:initial}(d). This results in the first three variants being valid, while the fourth has an invalid connection. FMCC constructs the partial circuit $PC(TX_0)$ by exploring all the partial circuits in $PC(S_0)$ and all the variants in $V(TX_0)$. 


\begin{figure}
    \centering
    \includegraphics[width=1\linewidth]{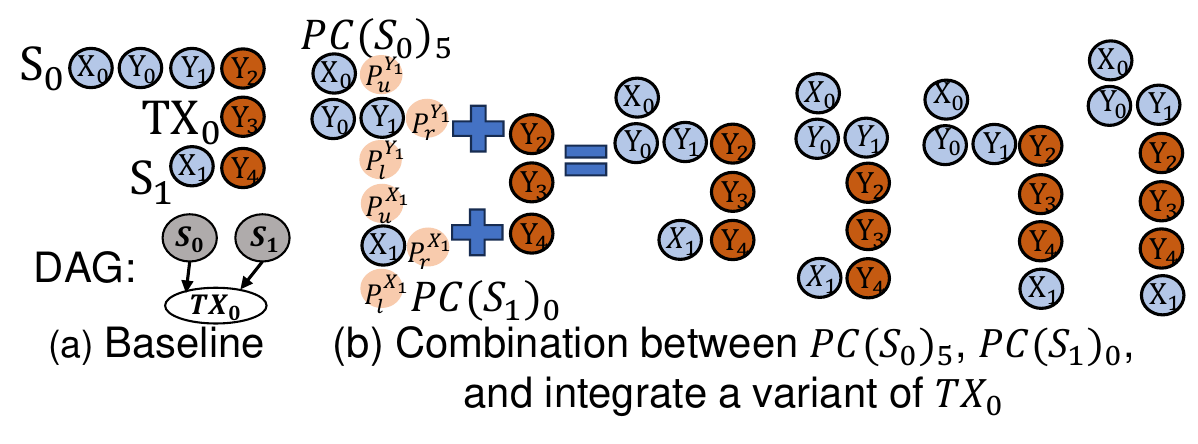}
    \caption{Example of partial circuits combination.}
    \label{fig:integrate}
\end{figure}

{\bf Partial circuit combination.} Selected components can have more than one parent, which additionally requires combination. For example, in Fig.~\ref{fig:integrate}(a), TX$_0$ has two parent components (S$_0$ and S$_1$). In this case, partial circuit combination of the two dependent sets $PC(S_0)$ and $PC(S_1)$ is required before TX$_0$ can be integrated. The combination follows an all-to-all manner. That is, each partial circuit from $PC(S_0)$ is combined with each partial circuit from $PC(S_1)$ in combination. Fig.~\ref{fig:integrate}(b) illustrates this using $PC(S_0)_5$ from $PC(S_0)$, the only partial circuit $PC(S_1)_0$ in $PC(S_1)$, and the first variant of TX$_0$ to generate four new partial circuits by exploring the anchor points $P = \{(P_r^{Y_1},P_r^{X1}), (P_l^{Y_1},P_r^{X1}), (P_r^{Y_1},P_u^{X1}), (P_l^{Y_1},P_u^{X1})\}$. Note that other anchor point combinations in $PC(S_0)_5$ and $PC(S_1)_0$ are not possible. Consequently, $PC(TX_0)$ is formed by combining the two sets of partial circuits and integrating the selected component while exploring all the anchor points.  



\begin{figure}
    \centering
    \includegraphics[width=1\linewidth]{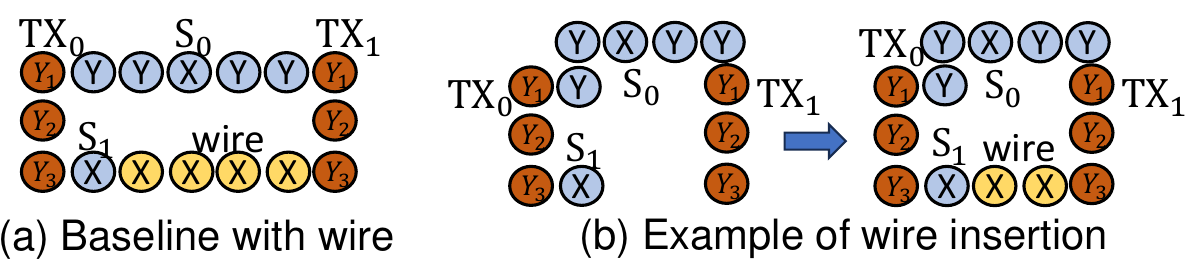}
    \caption{Example of wire insertion.}
    \label{fig:insertion}
\end{figure}

{\bf Wire insertion.} As discussed in Section~\ref{sec:cluster}, the function of a wire component is to connect a dependent component to its parent. Also, recall from Section~\ref{sec: opportu}, a wire can have valid variants with different even numbers X measurements. FMCC inserts wires on demand with appropriate wire variants to connect the dependent components whenever necessary. For example,  Fig.~\ref{fig:insertion}(a) shows a baseline MBQC circuit. Consider a valid partial circuit generated from the previous iteration in Fig.~\ref{fig:insertion}(b). FMCC inserts a wire (two X measurements) to connect TX$_1$ to its parent S$_1$.



\subsection{Partial Circuits Pruning}
\label{sec:pruning}
The partial circuit construction involves exploring all the variants and all the anchor points. It is difficult to track of all the partial circuits as the design space increases exponentially with the number of components. Fortunately, many of the partial circuits are invalid due to violation of the constraints (Section~\ref{sec:constraints}), such as the example in Fig.~\ref{fig:initial}(d) where there exists an invalid connection between X$_0$ and Y$_2$. FMCC drops all the invalid partial circuits. However, FMCC may still generate so many valid partial circuits that it is impractical to track them all. Therefore, to reduce the design space, FMCC employs a threshold $m$ to prune the set of partial circuits in each iteration. Specifically, for a partial circuit set $PC(c)$, we divide it into multiple sets where each set contains the partial circuits with the same width (i.e., $PC(c) = PC(c)^{w_1} \cup PC(c)^{w_2} \cup \cdots \cup PC(c)^{w_q}$, where $w_1$, $w_2$, $\cdots$, $w_q$ are the partial circuit width). For the partial circuit set of each width $PC(c)^{w_i}$, we retain the top $m$ partial circuits and drop the others to reduce the size of $PC(c)$. We choose different widths because it is more likely to produce a valid MBQC circuit with reduced depth. As discussed in Section~\ref{sec:motivate}, an MBQC circuit can reduce its depth by using more photon rows. Similarly, the depth of a partial circuit can be decreased by utilizing more rows (wider width). However, keeping only the partial circuits with a larger width may lead to an MBQC circuit width that exceeds the cluster state width after a combination. On the other hand, retaining partial circuits with a smaller width reduces this risk but may result in a deeper MBQC circuit. As a result, we retain $m$ partial circuits in $PC(c)^{w_i}$ for each width ${w_i}$ to achieve a valid and shorter MBQC circuit.


\begin{figure}
    \centerline{\includegraphics[width=\linewidth]{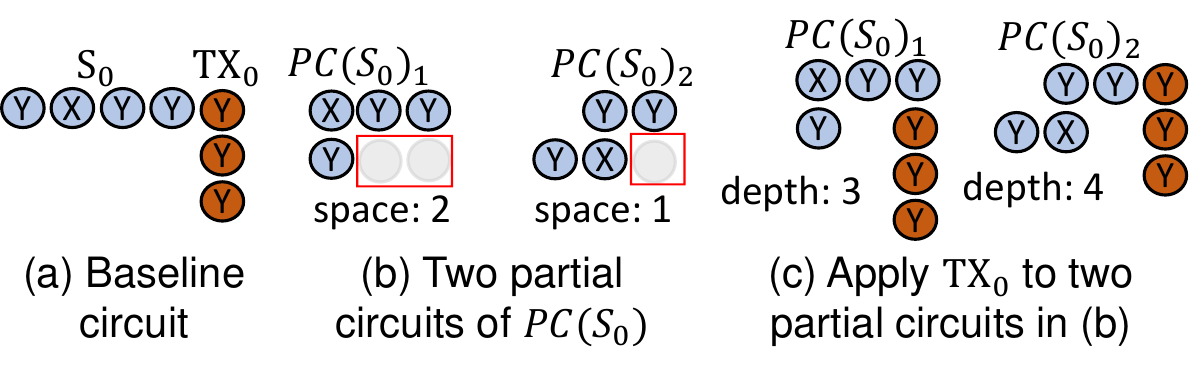}}
    \caption{Example of leveraging space.}
    \label{fig:space}
\end{figure}


We next answer the question: {\it which $m$ partial circuits to retain?} We leverage two metrics: i) {\it depth} and ii) {\it space}. 
Since the circuit depth is the objective of FMCC, a partial circuit with a shorter depth is preferred as it is more likely to generate a shorter MBQC circuit. If multiple partial circuits have the same width and depth, we use space to determine which $m$ partial circuits to retain. Particularly, space is defined as the number of unused photons after each rightmost measurement in a partial circuit. For example, Fig.~\ref{fig:space}(a) shows a circuit with two components S$_0$ and TX$_0$. The space is annotated in Fig.~\ref{fig:space}(b) for two partial circuits of $PC(S_0)$. The unused photons in the space may be used by component integration in the next iteration. Therefore, a larger space indicates a larger potential to generate short depth when integrating the next component. To illustrate this, consider the example in Fig.~\ref{fig:space}(c). If we choose the $PC(S_0)_1$ which has 2 spaces, we can generate the partial circuit with depth 3 after integrating TX$_0$. In contrast, if we choose $PC(S_0)_2$, which has 1 space, we will have the partial circuit with depth 4 after integrating TX$_0$. In FMCC, the width, depth, and space are tracked using a table where each row of the table is a partial circuit.

\begin{algorithm}
 \caption{FMCC}
 \label{algorithm}
 \let\COMMENT\undefined
 \begin{algorithmic}[1]
 \scriptsize
 \renewcommand{\algorithmicrequire}{\textbf{Input:}}
 \renewcommand{\algorithmicensure}{\textbf{Output:}}
 \REQUIRE Component DAG $G$, a cluster state $CS$, $m$
 \ENSURE Minimized-depth MBQC circuit
  \\ \textit{Initialization}: Partial circuits sets $PC$ = []
  \WHILE{$G$ != []}
  \STATE $c = independent(G)$ \label{algo:line2}
  \STATE $G.remove(c)$
  \STATE $V(c) = Generate\_variants(c)$
  \STATE $c_{parent1}, c_{parent2} = Find\_parent(c)$
  \STATE $PC(c)$ = []
  \FOR{$V_y$ in $V(c)$}
  \IF{$c_{parent2} == \emptyset$}
  \FOR{$PC_x$ in $PC(c_{parent1})$}
  \STATE $P = Find\_Anchor\_Points(PC_x)$
    \FOR{$P_1$ in $P$}
    \STATE $PC(c)_{x} = integrate(V_y, PC_x, P_1)$
    \STATE $PC(c)_{x} = insert\_wire(PC(c)_{x})$
    \STATE $PC(c).append(valid(PC(c)_{x}))$
    \ENDFOR
    \ENDFOR
  \ELSIF{$C_{parent2} != \emptyset$}
    \FOR{$PC_{x1}$ in $PC(c_{parent1})$}
    \FOR{$PC_{x2}$ in $PC(c_{parent2})$}
    \STATE $P^{c_{parent1}} = Find\_Anchor\_Points(PC_{x1})$
    \STATE $P^{c_{parent2}} = Find\_Anchor\_Points(PC_{x2})$
    \FOR{$P_1$ in $P^{c_{parent1}}$}
    \FOR{$P_2$ in $P^{c_{parent2}}$}
    \STATE $PC(c)_{x} = combine(V_y, PC_{x1}, PC_{X2}, P_1, P_2)$
    \STATE $PC(c)_{x} = insert\_wire(PC(c)_{x})$
    \STATE $PC(c).append(valid(PC(c)_{x}))$
    \ENDFOR
    \ENDFOR
    \ENDFOR
    \ENDFOR
  \ENDIF
  \ENDFOR
  \STATE $Table = create\_table(PC(c))$
  \STATE $PC = PC + select\_top\_m(PC(c), Table, m, CS))$
  \ENDWHILE
 \RETURN $shortest(PC[-1])$
 \end{algorithmic} 
 \end{algorithm}

\subsection{Detailed Algorithm of FMCC}
\label{sec: algo}

 \begin{figure*}
    \centerline{\includegraphics[width=\linewidth]{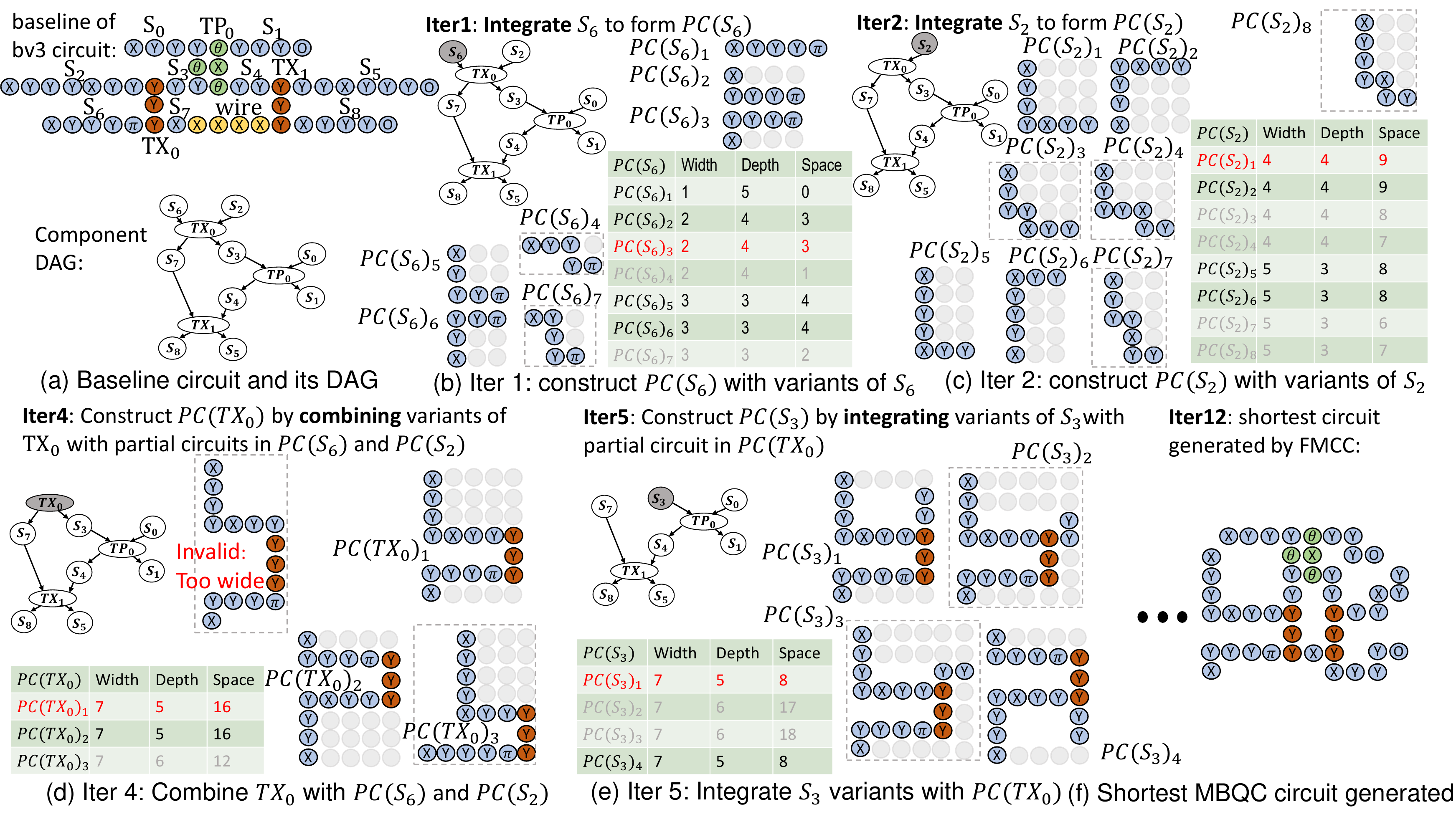}}
    \caption{Example of FMCC Constructing circuit variant.}
    \label{fig:design}
\end{figure*}


The detailed algorithm of FMCC is shown in Algorithm~\ref{algorithm}. The input is the component DAG $G$, a cluster state, and the parameter $m$ for retaining partial circuits. The output is the MBQC circuit with minimal depth. In each iteration, FMCC selects and removes an independent component $c$ from the component DAG (lines 2-3). FMCC then generates all its variants as a variant set $V(c)$ (line 4) and identifies its parent components (line 5). If $c$ has less than two parent components, FMCC conducts component integration that integrates each variant from $V(c)$ with each partial circuit $PC_x$ in $PC(c_{parent1})$ where $c_{parent1}$ is the parent component of $c$,  
while considering all anchor points $P$ (lines 8-14). When $c$ has two parent components, FMCC performs partial circuit combination of $PC(c_{parent1})$ and $PC(c_{parent2})$, and then integrates variants in $V(c)$ while exploring all the anchor points (lines 15-24). In this process, for each $PC(c)_x$ in $PC(c)$, FMCC inserts a wire if necessary (lines 13 and 23) and removes invalid $PC(c)_x$ (lines 14 and 24). Also, FMCC maintains a table to record the width, depth, and space for each partial circuit $PC(c)_x$ in $PC(c)$ (line 25), and uses the table to retain the $m$ partial circuits with the same width (line 26).


\subsection{A Walk-Through Example}
\label{sec:DP-single}
We now go through a real benchmark (i.e., 3-qubit Bernstein–Vazirani --- {\tt bv3}) to illustrate the steps of FMCC in Fig.~\ref{fig:design}. The baseline MBQC circuit of {\tt bv3} and its component DAG are shown in Fig.~\ref{fig:design}(a). Let us assume that the cluster state width $w$ is 8 and the retained partial circuit count $m$ is 2. There are a total of 12 iterations due to 12 non-wire components in the DAG. Due to the lack of space, Fig.~\ref{fig:design} only shows five iterations, which are sufficient to illustrate the steps of FMCC. In each iteration, we plot the table of the width, depth, and space of each partial circuit. We also highlight the chosen component in each iteration and show the remaining components in the DAG. For each partial circuit, photons contributing to space are indicated with gray-shaded circles.  

Initially, there are three independent components: S$_6$, S$_2$, and S$_0$. Thus, in the first three iterations, FMCC conducts the component integration for each independent component and generates $PC(S_6)$, $PC(S_2)$, and $PC(S_0)$. Fig.~\ref{fig:design}(b) shows the first iteration where S$_6$ is selected to construct $PC(S_6)$. There are more than 30 variants of S$_6$ and we show representative seven of them in Fig.~\ref{fig:design}(b), labeled from $PC(S_6)_1$ to $PC(S_6)_7$. Among the seven partial circuits, one has a width of 1, three have a width of 2, and three have a width of 3. Since $m$ is 2, FMCC will retain the only variant with width 1, two out of three variants with width 2, and two out of three variants with width 3. For $PC(S_6)_2$, $PC(S_6)_3$, and $PC(S_6)_4$ that have the same width 2, they also have the same depth. Therefore, the two with large spaces are retained, and the one with smaller space ($PC(S_6)_4$) is pruned (shaded in the table). Similarly,  For $PC(S_6)_5$, $PC(S_6)_6$, and $PC(S_6)_7$ that have the same width of 3, they also have the same depth. $PC(S_6)_7$ is pruned as it has the smallest space. In iteration 2, the procedure is similar. We show eight representative partial circuits of $PC(S_2)$ ($PC(S_2)_1$ to $PC(S_2)_8$) in Fig.~\ref{fig:design}(c). 
Among the eight partial circuits, $PC(S_2)_1$ to $PC(S_2)_4$ have the same width of 4 and the same depth of 4. $PC(S_2)_3$ and $PC(S_2)_4$ are pruned due to their smaller space. Similarly, $PC(S_2)_5$ to $PC(S_2)_8$ have the same width of 5 and the same depth of 3. $PC(S_2)_7$ and $PC(S_2)_8$ are pruned. 
In iteration 3, FMCC performs the same procedure to construct $PC(S_0)$. 




In iteration 4, TX$_0$ is independent and selected. Since TX$_0$'s parent components are in two partial circuit sets $PC(S_6)$ and $PC(S_2)$, FMCC performs combination of $PC(S_6)$ and $PC(S_2)$, and integrates $V(TX_0)$ with the combined partial circuits. We plot four example partial circuits generated after TX$_0$ integration in Fig.~\ref{fig:design}(d). The first one has a width of 9 which is larger than the fixed cluster state width of 8. Therefore, FMCC prunes it from $PC(TX_0)$. The other three partial circuits $PC(TX_0)_1$,  $PC(TX_0)_2$, and $PC(TX_0)_3$ have the same width of 7. Since  $PC(TX_0)_1$ and $PC(TX_0)_2$ have a shorter depth (5) compared to $PC(TX_0)_3$'s depth (6), they are retained in $PC(TX_0)$, and $PC(TX_0)_3$ is pruned. 

In iteration 5, S$_3$ becomes independent and $V(S_3)$ is integrated with the partial circuits in $PC(TX_0)$. We show four generated partial circuits with the same width in Fig.~\ref{fig:design}(e).  $PC(S_3)_1$ and $PC(S_3)_4$ are kept due to their shorter depth of 5. For the remaining iterations, similar integration and combination processes are conducted by FMCC to generate sets of partial circuits. Fig.~\ref{fig:design}(f) shows the MBQC circuit with the minimized depth after iteration 12.

\begin{figure}
    \centering
    \includegraphics[width=0.9\linewidth]{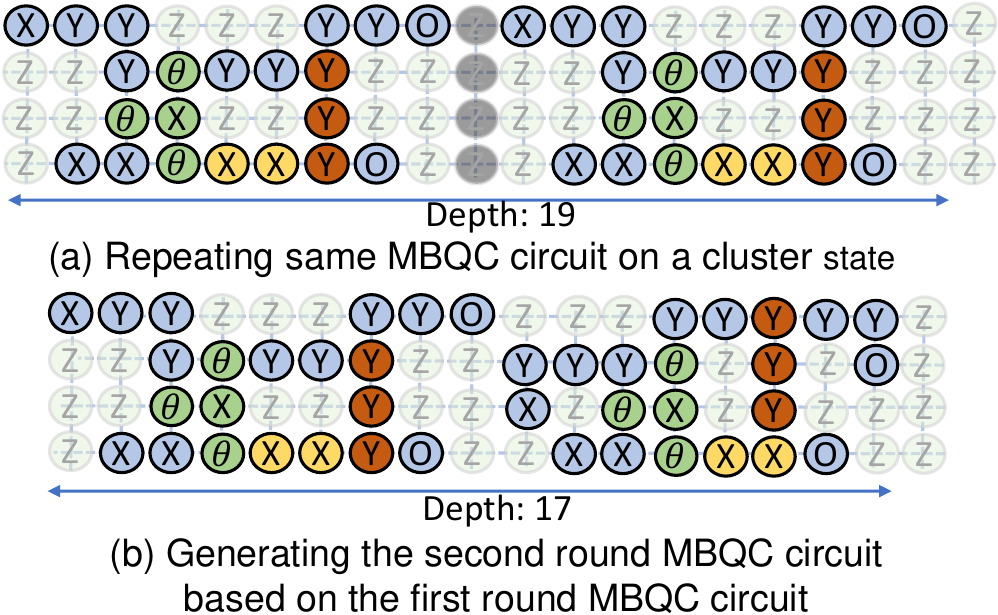}
    \caption{Executing two rounds of MBQC circuits.}
    \label{fig:continue}
\end{figure}

\subsection{Optimizing Multiple Rounds}
\label{sec: iterative}
The same quantum circuit usually executes multiple times to obtain stabilized output distribution~\cite{nielsen2001quantum}. This equivalently maps to the same MBQC circuits executing multiple times, which we referred to as ``rounds''. Fig.~\ref{fig:continue}(a) shows a quantum circuit that executes two rounds on a cluster state. As one can observe, a column of photons is cut out to separate the two rounds, annotated by gray photons, which extends the overall circuit depth. Fortunately, our proposed FMCC can be seamlessly applied to consecutive rounds of a quantum circuit. Specifically, for MBQC circuits with the same width of a round, FMCC keeps track of the top $m$ for that round and applies the same procedure (discussed in Section~\ref{sec: DAG} and Section~\ref{sec:pruning}) when picking up the components from the next round. This allows us to utilize the cut-out photons at the end of each MBQC round. For instance, Fig.~\ref{fig:continue}(b) shows a valid two-round execution where the second round does not have any connection with the first round, but is able to utilize the cut-out photons and reduce the depth to 17. It is also to be noted that, the second round has a different valid MBQC circuit compared to the first round.

\section{Evaluation}
\label{sec:eval}

\subsection{Experiment Setup}
{\bf Benchmarks and Metric: }We use five representative quantum applications (shown in Table~\ref{table:benchmark}) to evaluate FMCC: Quantum Fourier Transform ({\tt qft})~\cite{qft}, hardware efficient ansatz ({\tt hwea})~\cite{hwea}, Bernstein–Vazirani ({\tt bv})~\cite{bv}, instantaneous quantum polynomial time ({\tt iqp})~\cite{iqp}, and linear hydrogen atom chain ({\tt hc})~\cite{hc}. 
For each application, we evaluate three different numbers of qubits: 5 qubits (\textit{small}), 15 qubits (\textit{medium}), and 27 qubits (\textit{large}). Note that, {\tt hc} requires an even number of qubits~\cite{hc}. Therefore, we use 6, 14, and 26 qubits as the \textit{small}, \textit{medium}, and \textit{large} for {\tt hc}, respectively.
We use the cluster state depth reduction ratio as the metric to measure the effectiveness of our approach.

{\bf Baseline: }The baseline employs the state-of-the-art MBQC compilation on photonic cluster state~\cite{dac}. This baseline optimizes the depth of the MBQC circuit by leveraging independent gate reordering, redundant wire removal, and consecutive single-qubit gate combinations. We use this baseline to demonstrate the significant depth reduction brought by FMCC over the state-of-the-art.

{\bf Fixed cluster state width: }As discussed in Section~\ref{sec:mbqc basis}, the minimum cluster state width for a $N_q$-qubit benchmark is $2N_q-1$. We use two different fixed cluster state widths to evaluate FMCC: i) $1.25 \times (2N_q-1)$, referred to as \textit{C1}, and ii) $1.5 \times (2N_q-1)$, referred to as \textit{C2}.  

As discussed in Section~\ref{sec:pruning}, FMCC keeps track of the $m$ partial circuits for each width. For the main results, we empirically choose $m$ to 12, which we find is enough to provide the minimized depth. We also conduct a sensitivity study of different $m$ values in Section~\ref{sec: sensi}.

\begin{table}[h!]

  \centering
  \footnotesize
  \begin{tabular}{|c|c|c|c|c|c|c|c|c|c|}

    \hline
    \textbf{Appli-} & \multicolumn{3}{c|}{\textit{small}}  & \multicolumn{3}{c|}{\textit{medium}}  & \multicolumn{3}{c|}{\textit{large}}  \\ \cline{2-10}
    \textbf{cation} & \textbf{\textit{qubit}}& \textbf{\textit{C1}}& \textbf{\textit{C2}} & \textbf{\textit{qubit}}& \textbf{\textit{C1}}& \textbf{\textit{c2}}& \textbf{\textit{qubit}}& \textbf{\textit{C1}}& \textbf{\textit{C2}}\\
    \hline
    {\tt bv}  & 5& 12& 14 & 15& 37& 44 & 27& 67& 80\\
    \hline
    {\tt iqp}  & 5& 12& 14 & 15& 37& 44 & 27& 67& 80\\
    \hline
    {\tt hwea}  & 5& 12& 14& 15& 37 & 44 & 27& 67& 80\\
    \hline
    {\tt qft}  & 5& 12& 14 & 15 & 37& 44 & 27& 67& 80\\
    \hline
    {\tt hc}  & 6& 14& 17  & 14& 34& 41 & 26& 64& 77\\
    \hline
  \end{tabular}
  \caption{Number of qubits and width for each benchmark.}
  \label{table:benchmark}
\end{table}

\begin{figure}
    \centerline{\includegraphics[width=\linewidth]{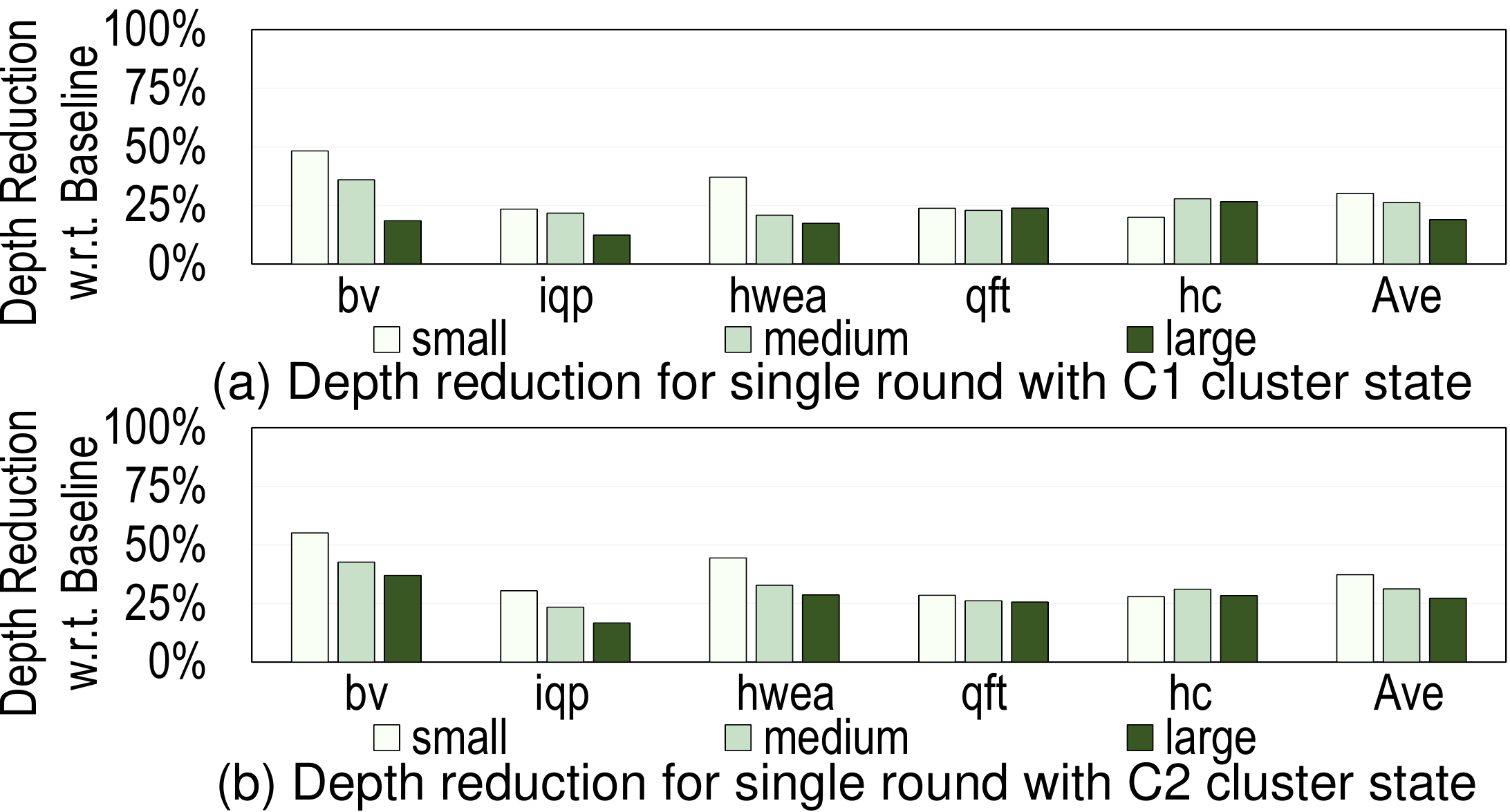}}
    \caption{Depth reduction for single round.}
    \label{fig:single round}
\end{figure}


\begin{figure}
    \centerline{\includegraphics[width=0.9\linewidth]{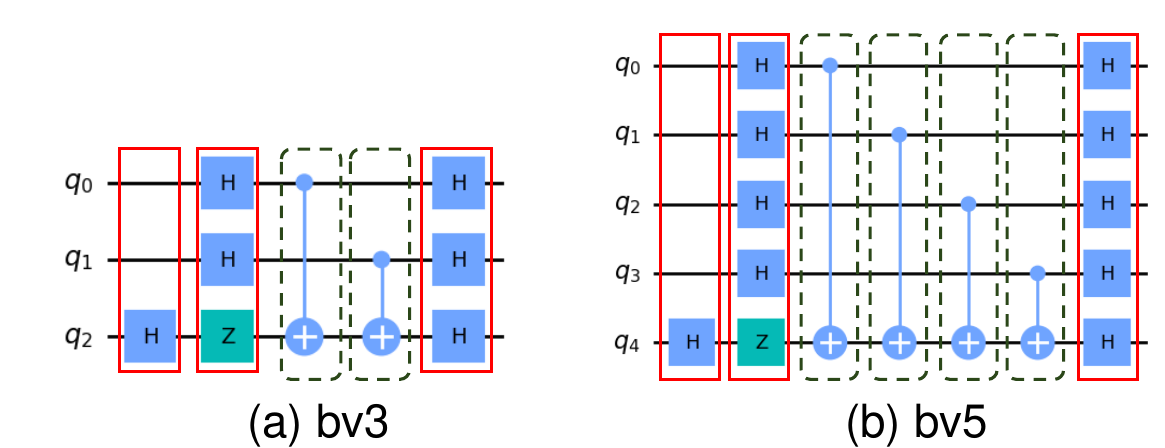}}
    \caption{Different qubit number {\tt bv} and the gate layers.}
    \label{fig:bv}
\end{figure}



\subsection{Single Round Results}
\label{sec:single}
We show the depth reduction ratios achieved by FMCC of a single round in Fig.~\ref{fig:single round} and multiple rounds (i.e., 100 rounds) in Fig.~\ref{fig:100 round}. Fig.~\ref{fig:single round} and Fig.~\ref{fig:100 round} also contain results under different fixed cluster widths (i.e., C1 and C2). 
As shown in Fig.~\ref{fig:single round}(a), FMCC achieves an average of 30.2\%, 26.2\%, and 19.0\% depth reductions for \textit{small}, \textit{medium}, and \textit{large} under \textit{C1} cluster state width, respectively. Similarly, under \textit{C2} width, FMCC achieves an average of 37.3\%, 31.3\%, and 27.0\% depth reductions, respectively. For all three numbers of qubits, the reductions increase from \textit{C1} to \textit{C2}, indicating that FMCC effectively leverages the redundant photons rows in wider cluster states to reduce the depth.

The depth reduction ratio decreases with more number of qubits for {\tt bv}, {\tt iqp}, and {\tt hwea}. To illustrate the reason, let us consider 3-qubit {\tt bv} in Fig.~\ref{fig:bv}(a) and 5-qubit {\tt bv} in Fig.~\ref{fig:bv}(b). In the figure, we annotate the layers of the single-qubit gate using red boxes and the layers of the two-qubit gate using green dash boxes. A layer is a collection of independent gates that can be performed concurrently. As one can observe, the number of two-qubit gate layers increases with the number of qubits. As such, there are more dependent TX components in the component DAG. For instance, Fig.~\ref{fig:bv}(a) has two layers of two-qubit gates, and Fig.~\ref{fig:bv}(b) has four layers of two-qubit gates. Recall that in Section~\ref{sec:motivate}, the TX and TP components have fewer variants and less reduction in depth compared to the S components. Therefore, with more TX components in large qubit counts, the depth reduction brought by FMCC also decreases. Similar results exist in the {\tt iqp} and {\tt hwea}, where with a larger qubit size, {\tt iqp} has more TP components and {\tt hwea} has more TX components.


In the {\tt hc}, the depth reduction increases from \textit{small} to \textit{medium} qubit count, and then decreases from \textit{medium} to \textit{large} qubit count. This is because {\tt hc} has long S components where many of its variants are invalid in the \textit{small} qubit count since it exceeds the fixed cluster state width (14 in \textit{C1} and 17 in \textit{C2}). In the \textit{medium} qubit count, the cluster state is wider, so more valid S components can be accommodated. However, in the \textit{large} qubit count, the depth reduction decreases because {\tt hc} also has an increased number of two-qubit gate layers as {\tt bv}, {\tt iqp}, and {\tt hwea}.  {\tt qft} exhibit similar depth reduction in all qubit counts. This is because the number of S components is proportionally increased with more qubits, providing a constant potential for depth reduction. 


\subsection{Multiple Round Results}
\label{sec:multi-round}
\begin{figure}
    \centerline{\includegraphics[width=\linewidth]{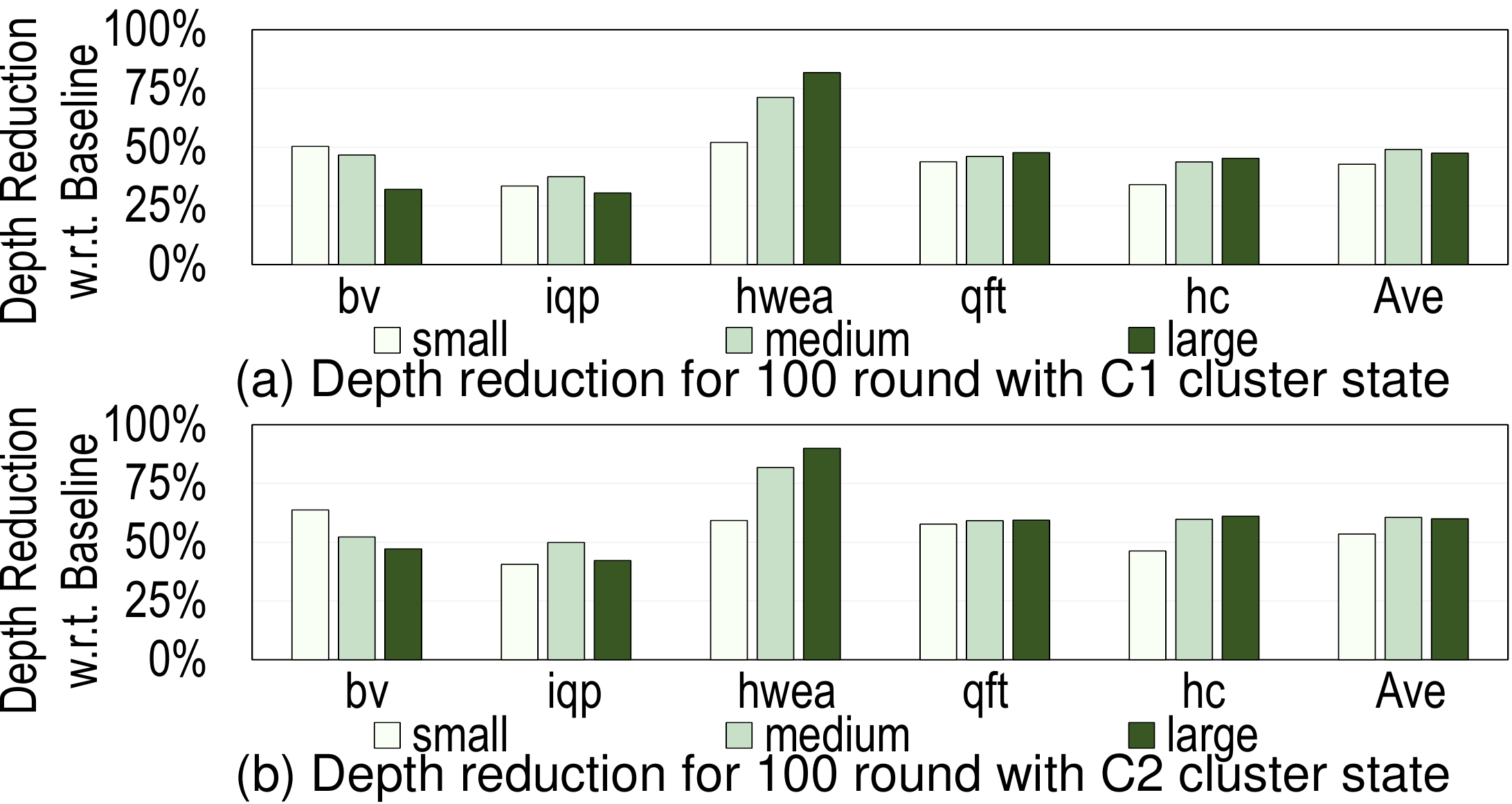}}
    \caption{Depth reduction for 100 rounds.}
    \label{fig:100 round}
\end{figure}

We evaluate one hundred rounds of each benchmark in Fig.~\ref{fig:100 round}. As one can observe, for \textit{C1}, FMCC achieves average of 42.7\%, 49.0\%, and 47.5\% depth reductions for \textit{small}, \textit{medium}, and \textit{large}, respectively. The depth reductions are 53.5\%, 60.6\%, and 60.0\% on \textit{C2}. Compared to the single-round results in Fig.~\ref{fig:single round}, FMCC achieves higher depth reduction, indicating its effectiveness in utilizing the cut-out photons between consecutive rounds as we discussed in Section~\ref{sec: iterative}. 

For the {\tt bv}, {\tt qft}, and {\tt hc}, the trend of depth reduction across the three qubit counts is similar to the trend observed in the single round results (Fig.~\ref{fig:single round}), indicating that FMCC effectively exploits both single round depth reduction and multi-round depth reduction. However, {\tt iqp} and  {\tt hwea} are two exceptions. For {\tt iqp}, the depth reduction increases from \textit{small} to \textit{medium} qubit count due to the effectiveness of FMCC multi-round optimization. However, the depth reduction drops from \textit{medium} to \textit{large} qubit count due to the increasing number of two-qubit gate layers (the same as the {\tt iqp} single round case). In contrast, the depth reduction of {\tt hwea} increases with more qubits. This is because with more number of qubits, the cut-out photons at the end of a {\tt hwea} round (i.e., after the last component in a round is integrated) increases. Recall our discussion in Section~\ref{sec: iterative} where the cut-out photons can be leveraged by consecutive rounds. Therefore, with more cut-out photons in the large qubit count, the depth reduction is also larger. 

\subsection{Photon Utilization}
\label{ssec:putil}

\begin{figure}
    \centerline{\includegraphics[width=\linewidth]{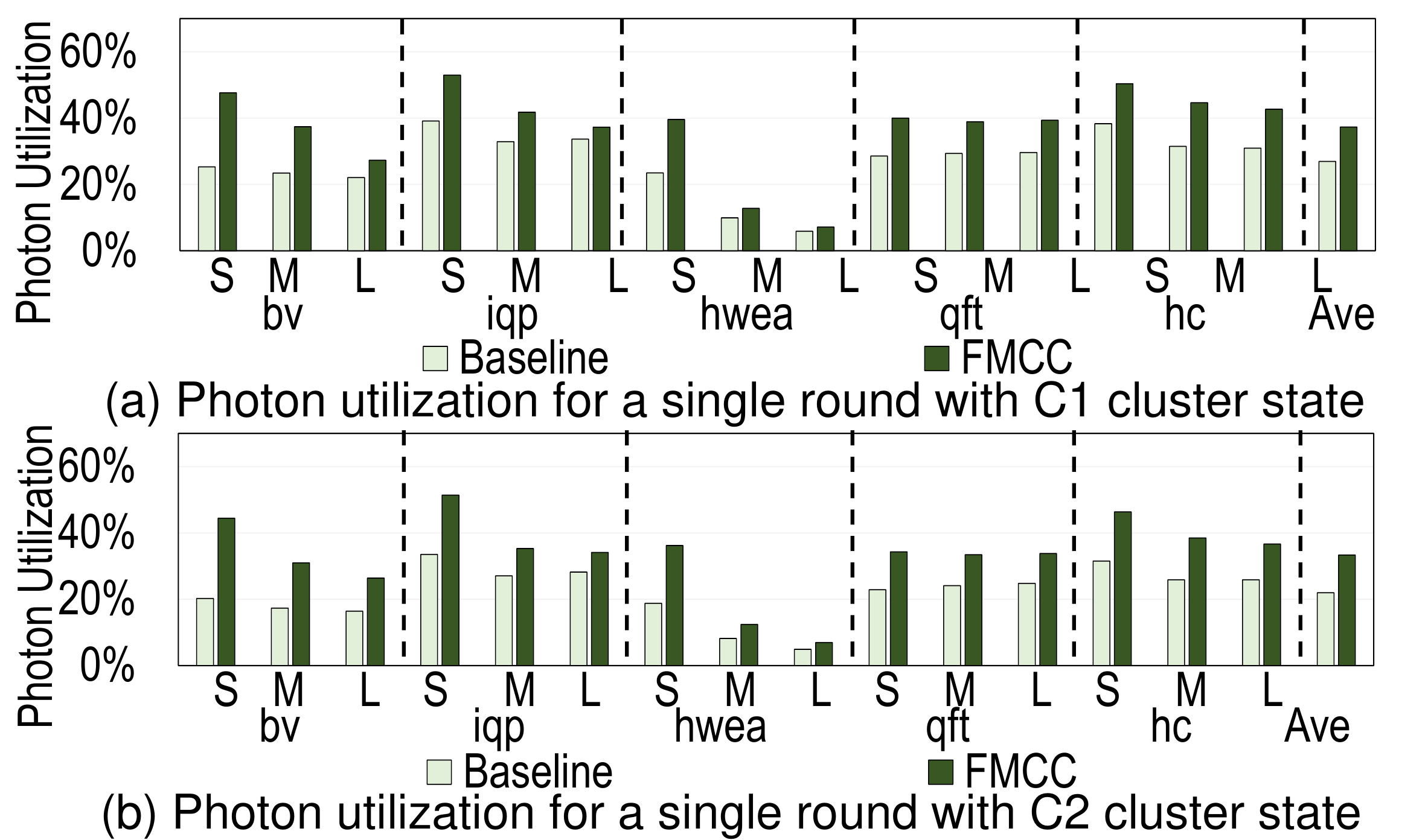}}
    \caption{Photon utilization for one round.}
    \label{fig:single utilization}
\end{figure}

\begin{figure}
    \centerline{\includegraphics[width=\linewidth]{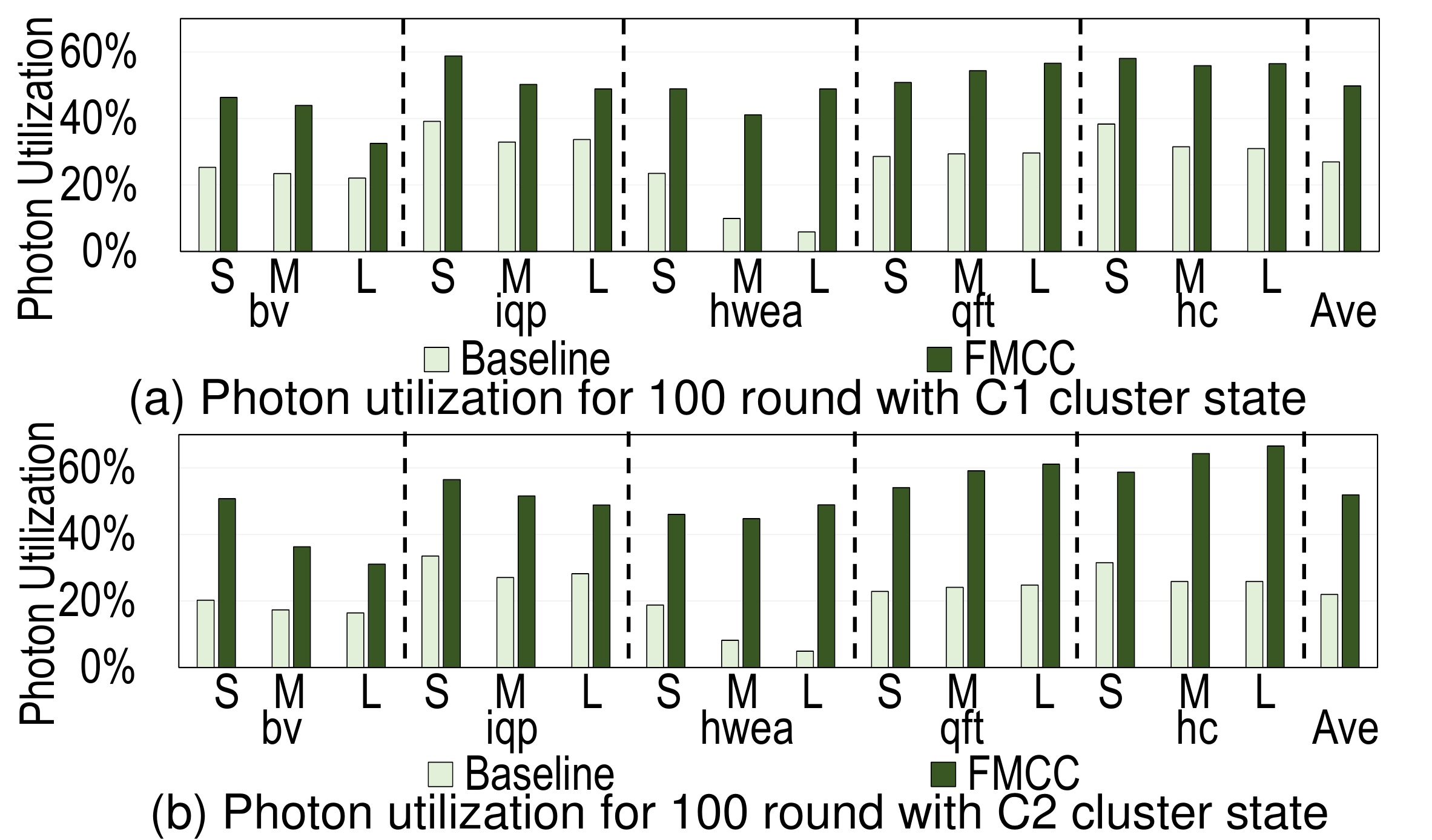}}
    \caption{Photon utilization for 100 rounds.}
    \label{fig:multi utilization}
\end{figure}

We plot the photon utilization of the single round in  Fig.~\ref{fig:single utilization} and the multi-round in Fig.~\ref{fig:multi utilization}. Photon utilization is defined as the ratio of non-z measurements in a cluster state to all the number of measurements in the cluster state. FMCC yields significant improvement in photon utilization compared to the baseline. On average, FMCC achieves the photon utilization of 37.3\%, 33.5\%, 50.1\%, and 51.8\% in single round C1, C2, and multi-round C1, C2, respectively. This is significantly higher compared to the baseline (27.0\%, 22.0\%, 27.0\%, and 22.0\% in single-round C1, C2, and multi-round C1, C2, respectively), indicating that FMCC effectively exploits the cutout photons to reduce the MBQC circuit depth.



\subsection{Sensitivity Study}
\label{sec: sensi}
\textbf{Large number of qubits.} We evaluate {\tt bv} and {\tt qft} using 50 and 100 qubits to study the scalability of FMCC using \textit{C2} fixed cluster state width (i.e., 149 rows for 50 qubits and 299 rows for 100 qubits). Overall, FMCC effectively reduces the circuit depth by 43.3\% and 59.8\% for {\tt bv-50} and {\tt qft-50}. The depth reductions are 42.1\% and 60.0\% for {\tt bv-100} and {\tt qft-100}. For {\tt bv}, the depth reduction ratio slightly drops with a larger qubit count due to the increased layers of the two-qubit gate (discussed in Section~\ref{sec:single}). In contrast, {\tt qft} has a similar depth reduction in 50 and 100 qubits due to the proportionally increased S components to the number of qubits~\cite{qft} (discussed in Section~\ref{sec:single}). 





\begin{figure}
    \centerline{\includegraphics[width=\linewidth]{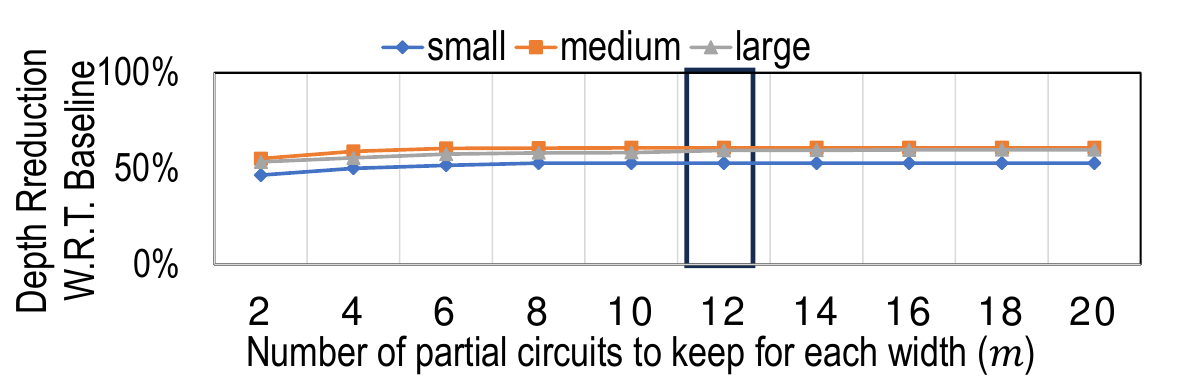}}
    \caption{Depth reduction with various of $m$.}
    \label{fig:sensitivity}
\end{figure}

\textbf{Number of partial circuits ($m$).}
Recall that we empirically set the number of partial circuits (i.e., $m$) to 12 for each cluster width in all previously reported results.  A large $m$ enhances the likelihood of finding a shorter cluster state depth, whereas a smaller $m$ has a lower searching cost due to the pruned search space. Fig.~\ref{fig:sensitivity} shows the averaged circuit depth reduction ratio when varying $m$ from 2 to 20 for {\it small}, {\it medium}, and {\it large} qubit counts. One can observe that the depth reduction ratios saturated at 8 for the \textit{small} and \textit{medium}, while for the \textit{large}, it saturated at 12.


\section{Related Works}
\label{sec:related}



Super-conducting quantum computation, one of the main paradigms in quantum computation has exhibited great generality and performance in many fields but is still limited by the physical device size and low coherence time. Multiple compilation frameworks ~\cite{2qan,scm,ccmqaoa} have been proposed to tackle these challenges and improve fidelity. For example, to address the problem that three-qubit gates are not natively supported at the physical level, gate decomposition~\cite{trios} and qubits compression~\cite{dqw} are applied to reduce the overhead introduced when mapping three-qubit gates onto physical devices. Also, to deploy a large-scale quantum circuit onto physical devices with limited size, Qompress~\cite{qompress} compresses two qubits into one four-state ququart and efficiently routes qubits to reduce communication and extra execution time brought by ququart gates. Besides, CaQR~\cite{caqr} leverages mid-circuit measurements to reduce qubit usages and the insertion of swap gates, thus leading to higher efficiency in resource utilization. What’s more, because the limited connectivity among physical qubits on super-conducting devices degrades the performance and fidelity of given circuits, SABRE~\cite{gushu} proposed a SWAP-based algorithm to search for qubit mapping solution to eliminate unnecessary operations, leading to higher efficiency and fidelity of the given circuit. However, the supported quantum gate of the quantum device and the connectivity among physical qubits are no longer concerns in MBQC. Meanwhile, there is a vast difference between the two quantum computation paradigms, thus this previous work cannot be applied to optimize MBQC circuits.


On the other hand, some pioneering steps are taken for leveraging the compilation to improve the fidelity or performance of MBQC. For MBQC based on the graph state, a multi-qubit state represented by a graph with a flexible geometry, it is scaled up by connecting resource states~\cite{rs} with highly error-prone quantum operations named fusion~\cite{fusion}. To reduce fusions and improve fidelity, several prior works build up the complete compilation flow~\cite{gs1, gs2, oneq} and apply the qubit delaying and routing algorithm to eliminate extra fusions. However, these works only focus on fusion reduction while our work targets to reduce the depth of a cluster state that is generated by a variety of techniques~\cite{coupleemitter1, coupleemitter2, pichler2017universal, 2023deterministic, quantumdot},  they are not applicable for reducing cluster state depth.


\section{Conclusion}  In this paper, we propose FMCC, which is flexible MBQC computation on cluster state with the goal of minimizing the cluster state depth. The proposed FMCC leverages both intra-component flexibility and inter-component flexibility to effectively utilize the redundant photons and rows in a fixed-width cluster state. Specifically, FMCC employs a dynamic programming approach to reduce the huge design space searching while avoiding sub-optimal results using simple heuristics. Experimental results using five representative quantum applications have shown that, compared to the state-of-the-art MBQC compilation, FMCC achieves 53.6\%, 60.6\%, and 60.0\% average depth reductions in small, medium, and large qubit counts, respectively.


\end{document}

%% file: macros.tex
\newcommand{\COMMENT}[1]{#1}
